\documentclass[year=23]{fmcad}

\usepackage{cite}
\usepackage{amsmath,amssymb,amsfonts}
\usepackage{algorithmic}
\usepackage{graphicx}
\graphicspath{ {images/} }
\usepackage{verbatim}
\usepackage[capitalise]{cleveref}
\usepackage{listings}
\usepackage{textcomp}
\usepackage{url}
\def\BibTeX{{\rm B\kern-.05em{\sc i\kern-.025em b}\kern-.08em
    T\kern-.1667em\lower.7ex\hbox{E}\kern-.125emX}}
\usepackage{booktabs}
\usepackage{multirow}
\usepackage{subcaption}

\usepackage{listing}
\usepackage{xspace}

\newcommand{\prettylstbtor}[0]{
  \lstset{
    frameround=fttt,
    frame=ltrb,
    breaklines=true,breakatwhitespace=true,
    basicstyle=\scriptsize\ttfamily, 
    showlines=true,
    showspaces=false,
    showtabs=false,
    escapeinside={~~},
    emphstyle=\sf\bf\color{blue},
    keywords={sort,bitvec,const,slice,const,state,init,next,eq,bad,
    one, zero, add, ones, mul, constd, ugt, and},
    keywordstyle={\color{blue}\bfseries}
}}

\newcommand{\prettylstverilog}[0]{
  \lstset{
    frameround=fttt,
    frame=ltrb,
    breaklines=true,breakatwhitespace=true,
    basicstyle=\scriptsize\ttfamily, 
    showlines=true,
    showspaces=false,
    showtabs=false,
    escapeinside={~~},
    emphstyle=\sf\bf\color{blue},
    keywords={module, input, output, reg, posedge, begin, if, else,
    endmodule, always, end, assert},
    keywordstyle={\color{blue}\bfseries}
}}

\newcommand{\prettylstllvm}[0]{
   \lstset{
    basicstyle={\scriptsize\ttfamily}, 
    keywordstyle={\color{blue}\bfseries},
    language=llvm,
    frameround=fttt,
    frame=ltrb,
    breaklines=true,breakatwhitespace=true,
}}

\newcommand{\prettylsttablegen}[0]{
   \lstset{
    basicstyle={\scriptsize\ttfamily}, 
    keywordstyle={\color{blue}\bfseries},
    frameround=fttt,
    frame=ltrb,
    keywords={Dialect, def, let, class, string, list, Commutative, SameOperandsAndResultType},
    breaklines=true,breakatwhitespace=true,
showlines=true,
   emph={},
   emphstyle=\sf\color{brown},
}}

\newcommand{\prettylstcpp}[0]{
   \lstset{
    basicstyle={\scriptsize\ttfamily}, 
    keywordstyle={\color{blue}\bfseries},
    language=c++,
    frameround=fttt,
    frame=ltrb,
    breaklines=true,breakatwhitespace=true,
}}

\newcommand{\llvm}[0]{\textsc{LLVM}\xspace}

\newcommand{\libfuzzer}[0]{\textsc{LibFuzzer}\xspace}
\newcommand{\klee}[0]{\textsc{Klee}\xspace}
\newcommand{\zthree}[0]{\textsc{Z3}\xspace}
\newcommand{\llvmir}[0]{\textsc{LLVM-IR}\xspace}
\newcommand{\mlir}[0]{\textsc{MLIR}\xspace}
\newcommand{\btor}[0]{\textsc{Btor2}\xspace}
\newcommand{\btormlir}[0]{\textsc{Btor2MLIR}\xspace}
\newcommand{\btordialect}[0]{\textsc{Btor Dialect}\xspace}
\newcommand{\smtlib}[0]{\textsc{Smt-Lib}\xspace}
\newcommand{\std}[0]{\textsc{Standard}\xspace}
\newcommand{\boolector}[0]{\textsc{Boolector}\xspace}
\newcommand{\btormc}[0]{\textsc{BtorMC}\xspace}
\newcommand{\btortools}[0]{\textsc{Btor2Tools}\xspace}
\newcommand{\smt}[0]{\textsc{Smt}\xspace}
\newcommand{\builtin}[0]{\textsc{Builtin}\xspace}
\newcommand{\yosys}[0]{\textsc{Yosys}\xspace}
\newcommand{\verilog}[0]{\textsc{Verilog}\xspace}
\newcommand{\seahorn}[0]{\textsc{SeaHorn}\xspace}

\newcommand{\tablegen}[0]{\textsc{TableGen}\xspace}

\begin{document}

\title{\textsc{Btor2MLIR}: A Format and Toolchain for Hardware Verification}

\author{\IEEEauthorblockN{Joseph Tafese \orcid{0000-0002-4062-0592}}
\IEEEauthorblockA{\textit{University of Waterloo}\\
Waterloo, Canada \\
jetafese@uwaterloo.ca  }
\and
\IEEEauthorblockN{Isabel Garcia-Contreras \orcid{0000-0001-6098-3895}}
\IEEEauthorblockA{\textit{University of Waterloo}\\
Waterloo, Canada \\
igarciac@uwaterloo.ca 
}
\and
\IEEEauthorblockN{Arie Gurfinkel \orcid{0000-0002-5964-6792}}
\IEEEauthorblockA{\textit{University of Waterloo}\\
Waterloo, Canada \\
agurfink@uwaterloo.ca }
}

\maketitle
 \pagestyle{plain}

\begin{abstract}
Formats for representing and manipulating verification problems are extremely important for supporting the ecosystem of tools, developers, and practitioners. A good format allows representing many different types of problems, has a strong toolchain for manipulating and translating problems, and can grow with the community. In the world of hardware verification, and, specifically, the Hardware Model Checking Competition (HWMCC), the \btor format has emerged as the dominating format. It is supported by \btortools, verification tools, and Verilog design tools like Yosys. In this paper, we present an alternative format and toolchain, called \textsc{Btor2MLIR}, based on the recent MLIR framework. The advantage of \textsc{Btor2MLIR} is in reusing existing components from a mature compiler infrastructure, including parsers, text and binary formats, converters to a variety of intermediate representations, and executable semantics of LLVM. We hope that the format and our tooling will lead to rapid prototyping of verification and related tools for hardware verification.  
\end{abstract}

\section{Introduction}
\label{sec:intro}
Hardware Verification has been one of the biggest drivers of formal verification 
research~\cite{HV}, with a history that spans many breakthroughs. The developments
in this field have thrived through organized events such as the Hardware Model Checking 
Competition (HWMCC)~\cite{HWMCC} which has run since~$2011$. \btor~\cite{BTOR} has 
emerged as the dominating format in this competition. \btor has been translated into 
several languages, for example, Constrained Horn Clauses (CHCs)%
\footnote{\url{https://github.com/zhanghongce/HWMCC19-in-CHC}}%
\footnote{\url{https://github.com/stepwise-alan/btor2chc}}
and 
\llvmir\footnote{\url{https://github.com/stepwise-alan/btor2llvm}} to make use 
of existing verification techniques. Universality, however, was not an objective 
of these projects, and thus, for these translations, be it to CHCs or to \llvmir, 
similar tasks had to be replicated.

During the past decade, the \llvm project~\cite{LLVM} has dedicated significant effort to universality. One such effort is \mlir~\cite{MLIR}, a project that proposes a generic intermediate representation with operations and types common to many programming languages. \mlir was designed to be easily extensible, by providing tools to build new intermediate representations (IR) as dialects of the base \mlir. This eases the creation of new compilers, circumventing the need to re-implement core technologies and optimizations. Extensibility and scalability are what \mlir strives for, making it a great candidate for the creation of new tools and formats that represent many types of problems and have strong tool support for manipulating and translating problems.

During the same time, with the rise of \llvm  as a compiler infrastructure,
many software verification tools have been built for \llvmir programs.
Existing tools tackle this hard problem in many ways. For example, dynamic verification 
is implemented in \libfuzzer~\cite{LibFuzzer}, a fuzzer, and \klee~\cite{KLEE},
a symbolic execution engine;
SMT-based static verification is implemented in \seahorn~\cite{SeaHorn} both as 
Bounded and Unbounded Model Checking; and \textsc{Clam}~\cite{DBLP:conf/vstte/GurfinkelN21} 
static analysis that analyzes \llvmir statically using abstract interpretation.

This paper contributes \btormlir, a format and toolchain for hardware verification. It is built on \mlir to incorporate advances and best practices in compiler infrastructure, compiler design, and the maturity of \llvm. At its core, \btormlir provides an intermediate representation for \btor as an \mlir dialect. This dialect has an encoding very close to \btor and preserves \btor's semantics. This design not only facilitates the creation of a new format for hardware verification but also simplifies the extension of this format to support future targets by using \mlir for all intermediate representations. For example, \btormlir can be used to generate \llvmir from our custom \mlir dialect. The value of this approach is quite evident in CIRCT~\cite{CIRCT}, an open-source project, that applies this design to tackle the inconsistency and usability concerns that plague tools in the Electronic Design Automation industry. Although it has a different goal than \btormlir, both projects draw great benefit from adapting the benefits of an \mlir design to their respective fields. 

As an added bonus, using \btormlir to generate \llvmir enables the reuse of established 
tools to apply software verification techniques to verify hardware circuits. 
To illustrate the usability of the toolchain, a new model checker is developed using \seahorn. The results are 
compared to \btormc\cite{BTOR}, a hardware model checker provided 
by the creators of \btor.

The rest of the paper is organized as follows. \cref{sec:background} lays some background. Our format and toolchain, \btormlir, is described in \cref{sec:btormlir}. We discuss its correctness in \cref{sec:correctness} and evaluate the tool in \cref{sec:evaluation}. We close with a note on related works in \cref{sec:related} and conclude in \cref{sec:conclusion}.

\section{Background}
\label{sec:background}

\begin{figure*}[t]
\centering
\begin{subfigure}[b]{0.31\textwidth}
    \includegraphics[width=1\textwidth]{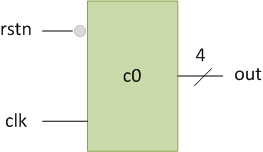}
    \vspace{0.2in}
    \caption{Circuit.}
    \label{fig:circuit-counter}
\end{subfigure}
\hfill%
\begin{subfigure}[b]{0.33\textwidth}
\prettylstverilog
\begin{lstlisting}
module counter (
   input clk, 
   output reg[3:0] out);
  always @ (posedge clk) 
  begin 
    assert (out != 15)
    out <= out + 1
  end
endmodule
\end{lstlisting}
\caption{Counter in \verilog.}
\label{fig:verilog-counter}
\end{subfigure}
\hspace*{\fill}%
\begin{subfigure}[b]{0.26\textwidth}
\prettylstbtor
\begin{lstlisting}
1 sort bitvec 4
2 zero 1
3 state 1 out
4 init 1 3 2
5 one 1
6 add 1 3 5
7 next 1 3 6
8 ones 1
9 sort bitvec 1
10 eq 9 3 8
11 bad 10
\end{lstlisting}
\caption{Counter in \btor.}
\label{fig:btor-counter}
\end{subfigure}
\caption[]{4-bit counter.}
\label{fig:counter}
\end{figure*}

\paragraph*{\btor} \btor \cite{BTOR} is a format for quantifier-free formulas and  
circuits over bitvectors and arrays, with \smtlib \cite{SMTLIB} 
semantics, that is used for hardware verification. \btor
files are often generated using tools like \yosys \cite{Yosys}, from the
original design in a language like \verilog \cite{Verilog}. A simple 4-bit
counter is shown in \cref{fig:counter}.  
Its corresponding description, in \verilog, is 
shown in \cref{fig:verilog-counter}. The circuit 
updates its output at each step starting from $0$ to its maximum value,
$15$. It also has the safety property that the output should 
not be equal to $15$, shown by the assertion in \cref{fig:verilog-counter}. The circuit together 
with the desired safety property
are captured in \btor in \cref{fig:btor-counter}.
First, a bitvector of width $4$ is defined as '\texttt{1}' in line~$1$. Sort are used later when declaring registers and operations. For example, lines $2$, $3$, $5$,
and $8$ refer to sort '\texttt{1}', respectively, by declaring '\texttt{2}' to be a zero bitvector ($0000$) (line 2),  state \texttt{out} to be a register of sort '\texttt{1}' (line 3), '\texttt{5}' to be a one bitvector ($0001$) (line 5) and '\texttt{8}' to be  bitvector of ones ($1111$) (line 8). On line~$4$,
\texttt{out} is initialized with value '\texttt{2}'. On line~$7$, the transition
function is defined (activated at each clock edge), by assigning the next state
of \texttt{out} to the value \texttt{out} incremented by one (the result of line
$6$). Finally, a safety property is defined
in line~$11$ with the keyword \texttt{bad}, requiring that the equality of line~$10$ does not hold. That is, the value of \texttt{out} is never $1111$. 
Note that no clock is specified in \cref{fig:btor-counter}.
In \btor it is always assumed that there is one single clock, and the keyword \texttt{next} is used to declare how registers are updated after a clock cycle. For a register that has not been assigned a next value, it will get a new non-deterministic value or keep it's initial value (if one was given).

\paragraph*{\btormc} \btormc~\cite{BTOR} is a bounded model checker (BMC) for \btor. \btormc
generates verification conditions as \smt formulas and uses \boolector
\cite{BTOR} as an \smt solver. Based on the satisfiability result of the
formula, \btormc on our example tells us that the safety property is violated, as expected,
since \texttt{out} does reach a state with value $1111$. 

\paragraph*{\mlir} Multi-Level Intermediate Representation (\mlir) 
\cite{MLIR} is a project that
was developed for TensorFlow~\cite{TensorFlow} to address
challenges faced by the compiler industry at large: modern languages
end up creating their own high-level intermediate representation (IR) and
the corresponding technologies. Furthermore, these domain-specific 
compilers have to be recreated for different compilation and optimization 
targets and do not easily share a common infrastructure or 
intermediate representations. To remedy this,
\mlir facilitates the design and implementation of code generators, translators,
and optimizers at different levels of abstraction and also across application
domains, hardware targets, and execution environments.

Modern languages vary in the set of operations and types that they use, hence 
the need to create domain-specific high-level IRs. \mlir addresses this problem 
by making it easy for a user to define their own dialects.
An \mlir dialect captures the operations and types of a target language. It is
created using \tablegen, a domain specific language for defining \mlir dialects. 
It is used to automatically generate code to manipulate the newly defined dialect 
including its Abstract Syntax Tree (AST) and parsing. 
\mlir tools and optimizations such as
static single assignment, constant propagation, and dead-code elimination can be 
applied off the shelf to custom \mlir dialects. These capabilities make \mlir a reusable
and extensible compiler infrastructure. One of its strengths is the builtin
dialects it introduces, such as a \builtin, \std, and \llvm dialects%
\footnote{
\url{https://github.com/llvm/llvm-project/tree/release/14.x/mlir/include/mlir/Dialect}},
among others. These dialects make it possible to have a rich infrastructure for
dialect conversion that enables a user to define pattern-based rewrites of
operations from one dialect to another. For example, a dialect conversion pass
is provided to convert operations in the \std dialect to operations in the \llvm
dialect. \mlir also provides an infrastructure for user-defined language
translation passes. One such pass that is provided out of the box is a
translation from \llvm dialect to \llvmir.

\section{\btormlir}
\label{sec:btormlir}

\begin{figure*}[t]
  \centering
  \includegraphics[width=1\textwidth]{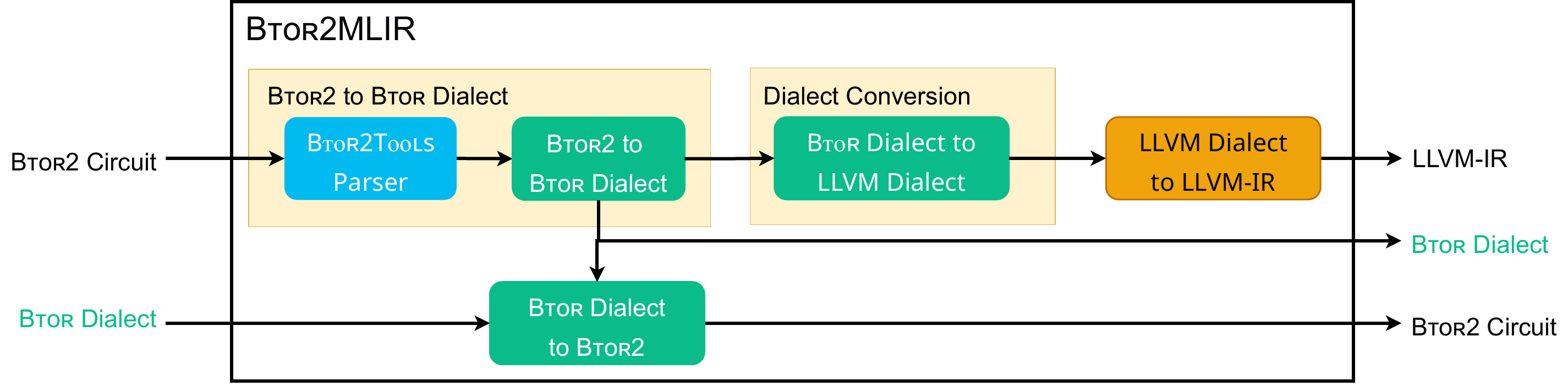}
  \caption{\btormlir Architecture.}
  \label{fig:architecture}
\end{figure*}

We present our tool, \btormlir, 
which  contributes the \btordialect, 
and three modules on the  existing \mlir infrastructure: a \btor to 
\btordialect translation pass, a \btordialect to \btor translation pass
and a dialect conversion pass from \btordialect to \llvm dialect.
Our tool has approximately $3\,900$ lines of C\texttt{++} code and $1\,200$ 
lines of \tablegen. 
\cref{fig:architecture} shows the architecture of our tools with
our contributions highlighted in green. \btormlir uses the original \btor 
parser provided in \btortools~\cite{BTOR}, marked in blue, and \mlir builtin passes, 
marked in brown. \btormlir is open-sourced and publicly available on GitHub%
\footnote{\url{https://github.com/jetafese/btor2mlir/tree/llvm-14}}.

We illustrate how each of the components of \btormlir works by translating 
a factorial circuit, shown in \cref{fig:btor-factorial}, that is described
in \btor. There are two safety properties, one per \texttt{bad} 
statement. Line~$14$ states that the loop counter, \texttt{i}, reaches $15$. 
Line~$19$ states that the value of \texttt{factorial} is always even. 

\begin{figure*}[t]
\centering
\vfill
\begin{subfigure}[b]{0.18\textwidth}
\prettylstbtor
\begin{lstlisting}
1 sort bitvec 4
2 one 1
3 state 1 factorial
4 state 1 i
5 init 1 3 2
6 init 1 4 2
7 add 1 4 2
8 mul 1 3 4
9 next 1 4 7
10 next 1 3 8
11 ones 1
12 sort bitvec 1
13 eq 12 4 11
14 bad 13
15 slice 12 3 0 0
16 constd 1 3
17 ugt 12 4 16
18 and 12 17 15
19 bad 18
\end{lstlisting}
\caption{Factorial in \btor.}
\label{fig:btor-factorial}
\end{subfigure}
\hspace{5mm}%
\begin{subfigure}[b]{0.32\textwidth}
\prettylstllvm
\begin{lstlisting}
module {
  func @main() {
    %0 = btor.constant 1 : i4
    br ^bb1(%0, %0 : i4, i4)
  ^bb1(%1: i4, %2: i4): 
    %3 = btor.constant 1 : i4
    %4 = btor.add %2, %3 : i4
    %5 = btor.mul %1, %2 : i4
    %6 = btor.constant -1 : i4
    %7 = btor.cmp eq, %2, %6 : i4
    btor.assert_not(%7)
    %8 = btor.constant 0 : i4
    %9 = btor.constant 0 : i4
    %10 = btor.slice %1, %8, %9 : i4, i1
    %11 = btor.constant 3 : i4
    %12 = btor.cmp ugt, %2, %11 : i4
    %13 = btor.and %12, %10 : i1
    btor.assert_not(%13)
    br ^bb1(%5, %4 : i4, i4)
  }
}
\end{lstlisting}
\caption{Factorial in \btordialect.}
\label{fig:btor-dialect}
\end{subfigure}
\hspace{5mm}%
\begin{subfigure}[b]{0.30\textwidth}
\prettylstllvm
\begin{lstlisting}
declare void @__VERIFIER_error()
define void @main() !dbg !3 {
  br label %1
1:      ; preds = %14, %0
  %2 = phi i4 [%5,%14], [1,%0]
  %3 = phi i4 [%4,%14], [1,%0]
  %4 = add i4 %3, 1
  %5 = mul i4 %2, %3
  %6 = icmp eq i4 %3, -1
  %7 = xor i1 %6, true
  br i1 %7, label %8, label %15
8:      ; preds = %1
  %9 = lshr i4 %2, 0
  %10 = trunc i4 %9 to i1
  %11 = icmp ugt i4 %3, 3
  %12 = and i1 %11, %10
  %13 = xor i1 %12, true
  br i1 %13, label %14, label %16
14:      ; preds = %8
  br label %1
15:      ; preds = %1
  call void @__VERIFIER_error()
  unreachable
16:      ; preds = %8
  call void @__VERIFIER_error()
  unreachable
}
\end{lstlisting}
\caption{Factorial in \llvmir.}
\label{fig:llvm-ir}
\end{subfigure}
\caption[]{\btor to \btordialect.}
\label{fig:translation}
\end{figure*}

\paragraph*{\btordialect}
Our first contribution is the \btordialect, an \mlir dialect to represent 
\btor circuits. \cref{fig:btor-dialect} shows the \btordialect code corresponding 
to \cref{fig:btor-factorial}. It represents the execution of the circuit using 
an \mlir function \texttt{main}. The control flow is explicit, 
using a standard \mlir representation of basic blocks with arguments and branches. 
The example has two basic blocks: an unnamed initial block (\texttt{bb0}) and 
a block \texttt{bb1}. Circuit initialization is modeled by instructions
in \texttt{bb0}, and each cycle by instructions in \texttt{bb1}. Note that 
\texttt{bb1} has two predecessors: \texttt{bb0} for initialization and \texttt{bb1} 
for each cyle. Bitvector types are mapped to integer types 
(provided by \mlir), for example, \texttt{bitvec 4} becomes \texttt{i4}. 
Each operation in the \btordialect, prefixed with \texttt{btor}, models a 
specific \btor operation. For example, \texttt{btor.mul} is \texttt{mul}, and 
\texttt{btor.slice} is \texttt{slice}. Safety properties such as 
\texttt{bad} are represented by \texttt{btor.assert\_not}. Special 
operators such as \texttt{one}, \texttt{ones} and \texttt{constd} 
are represented by the \texttt{btor.constant} operation with the expected 
integer value. Boolean operators are represented by \texttt{btor.cmp}. 
For example, \texttt{eq} becomes \texttt{btor.cmp eq}.

\paragraph*{Translating \btor to \btordialect} 
\btormlir takes \btor circuits as input, using \btortools to create a data structure 
for each \btor line. Our pass then generates a program in \btordialect by constructing the 
appropriate \mlir AST. Each \btor operator is mapped to a unique operation 
in \btordialect, a capability that is greatly simplified and enabled 
by the \mlir infrastructure. 

A program in \mlir can be written using multiple dialects since the 
\mlir framework enables the interaction of multiple IRs.
To enable this capability, \mlir provides dialects that are designed to serve 
as building blocks for more domain-specific dialects. We utilized the framework by 
building the \btordialect using the \std and \builtin dialects. For example, 
we use the \textit{module}, \textit{func} and \textit{bb} operations from 
\builtin. We utilize the \textit{br} operation in the \std dialect to enable 
interaction between the two basic blocks in \cref{fig:btor-dialect}. 
This approach is consistent with the intended use of the \std and \builtin
dialects. It saves time and effort since we do not need to recreate operations 
that already exist in other dialects. Furthermore, \mlir provides a conversion pass from 
\std dialect to \llvm dialect, making it worthwhile to build \btordialect
on top of the \builtin and \std dialects. 

\paragraph*{Dialect conversion} 
The \btormlir conversion pass from \btordialect to \llvm dialect utilizes the 
\mlir infrastructure for pattern-based rewrites. It rewrites \btordialect 
operations into \llvm dialect operations.
For most operations in \btordialect there exists a semantically equivalent 
operation in \llvm dialect. For example, \textit{btor.constant} in 
\cref{fig:btor-dialect} is converted to \textit{llvm.mlir.constant} in 
\llvm dialect. For some operations, an equivalent in  
\llvm dialect does not exist, in these cases it is required to rewrite them into several 
\llvm operations (e.g., in
\textit{btor.slice}) and/or to modify the module structure (e.g., \textit{btor.assert\_not}). 
In \llvm dialect, \textit{btor.slice} is replaced by a logical 
shift right, \textit{llvm.lshr}, and a truncation operation, 
\textit{llvm.trunc}. \textit{btor.assert\_not} is mapped to 
a new basic block in the \llvm dialect that has the \textit{llvm.unreachable} 
operation. We split the basic block \textit{bb1}, in \cref{fig:btor-dialect},
by adding a conditional branch, \textit{llvm.cond\_br},
to direct control flow to the new block when the assertion is satisfied.

\paragraph*{Translate \llvm Dialect to \llvmir}  \btormlir uses a translation pass from \llvm dialect to \llvmir, provided by  \mlir. Note the optimizations in the resulting \llvmir, shown in \cref{fig:llvm-ir}, such as constant propagation and phi nodes.

\section{Correctness}
\label{sec:correctness}

\begin{table}[t]
\setlength{\tabcolsep}{5pt}
    \begin{tabular}{lrcrcrcr}
        & \multicolumn{3}{c}{original} & & \multicolumn{3}{c}{roundtrip} \\ [1mm]
        \cline{2-4} \cline{6-8} \\ [-2mm]
         & time & safe/unsafe & TO & & time & safe/unsafe & TO \\ [1mm]
         \toprule
         \bf bitvectors & & & & & \\[2mm]
        \tt wolf/18D &  157 & 34/0 & 2 & & 168 & 34/0 & 2 \\
        \tt wolf/19A &  146 & 0/1 & 17 & & 151 & 0/1 & 17 \\
        \tt wolf/19B & 2 & 3/0 & 0 & & 2 & 3/0 & 0 \\
        \tt wolf/19C & 834 & 108/0 & 5 & & 797 & 108/0 & 5 \\
        \tt 19/beem & 278 & 9/2 & 4 & & 280 & {\bf 10}/2 & 3 \\
        \tt 19/goel & 190 & 26/2 & 43 & & 176 & 26/2 & 43 \\
        \tt 19/mann & 4\,442 & 29/15 & 9 & & 4\,751 & {\bf 30}/15 & 8 \\
        \tt 20/mann & 257 & 10/5 & 0 & & 268 & 10/5 & 0 \\[1mm]
        \toprule 
        \multicolumn{3}{l}{\bf bitvectors + arrays} & & & \\[2mm]
        \tt wolf/18A & 70 & 20/0 & 0 & & 71 & 20/0 & 0 \\
        \tt wolf/19B & 2 & 2/3 & 0 & & 2 & 2/3 & 0 \\
        \tt 19/mann & 126 & 1/1 & 1 & & 138 & 1/1 & 1 \\
        \tt 20/mann & 18 & 3/3 & 0 & & 18 & 3/3 & 0 \\[1mm]
        \bottomrule
        \vspace{0.1in}
    \end{tabular}
\caption{Comparing round tripped files.}
\label{tab:roundtrip}
\end{table}

When introducing a new tool or framework to the community, there is always a question of how polished it is. \btormlir builds on two mature frameworks: \btortools and \mlir. This is done not only because of the frameworks' functionalities, but because they have been extensively reviewed, used, and tested. 
\btortools has been widely used in the hardware model-checking community since its introduction in 2018. \mlir builds on \llvm, a compiler framework that has been used and improved over numerous projects in the last two decades and is actively supported by industry. 

Specifically, \btormlir uses the parser from \btortools to generate corresponding operations and functions in the \btordialect of \mlir. 
The \btordialect is written in \tablegen --- an \mlir domain-specific language for dialect creation. We show how our dialect and the class of binary operations are defined in \cref{fig:create-dialect}. For example, the \texttt{BtorBinaryOp} class defines a class of operations that have two arguments \texttt{lhs, rhs} and a result \texttt{res}. It also has a trait \texttt{SameOperandsAndResultType} to enforce that \texttt{lhs, rhs} and \texttt{res} have the same type. Finally, the class specifies how the default \mlir parsers and printers should handle such operations. We create our arithmetic operations as shown in \cref{fig:create-ops}. We mark relevant operations as \texttt{Commutative}. Operation descriptions are not shown for simplicity. We ensure that each \btor operator has a one-to-one mapping with an operation in the \btordialect so that the translation from \btor to \btordialect is lossless and preserves \btor semantics. 

\begin{figure*}[t]
\centering
\vfill
\begin{subfigure}[b]{0.57\textwidth}
\prettylsttablegen
\begin{lstlisting}
def Btor_Dialect : Dialect {
...
}

class BtorArithmeticOp<string mnemonic, list<Trait> traits = []> : Op<Btor_Dialect, mnemonic, traits>;


class BtorBinaryOp<string mnemonic, list<Trait> traits = []> :
  BtorArithmeticOp<mnemonic, !listconcat(traits [SameOperandsAndResultType])>,
  Arguments<(ins SignlessIntegerLike:$lhs, 
                SignlessIntegerLike:$rhs)>,
  Results<(outs SignlessIntegerLike:$result)> 
{
    let assemblyFormat = "$lhs `,` $rhs attr-dict `:` type($result)";
}
\end{lstlisting}
\caption{Creating \btordialect.}
\label{fig:create-dialect}
\end{subfigure}
\hspace{5mm}%
\hspace{5mm}%
\begin{subfigure}[b]{0.35\textwidth}
\prettylsttablegen
\begin{lstlisting}
def AddOp : BtorBinaryOp<"add", [Commutative]> {
    ...
}

def SubOp : BtorBinaryOp<"sub"> {
    ...
}

def MulOp : BtorBinaryOp<"mul", [Commutative]> {
    ...
}

def UDivOp : BtorBinaryOp<"udiv"> {
    ...
}
\end{lstlisting}
\caption{Creating Operations for \btordialect.}
\label{fig:create-ops}
\end{subfigure}
\caption[]{Using \tablegen for Dialect Creation.}
\label{fig:tablegen}
\end{figure*}

\btormlir relies on the optimization, folding, and canonization passes that \mlir provides in its translation from the \llvm Dialect in \mlir to \llvmir. \mlir also provides the mechanism for pattern-based rewrites which has helped us avoid the introduction of undefined behavior into the resulting \llvmir. We show an example of this in \cref{fig:cpp}. \mlir allows us to identify which operations in the \btordialect we want to replace at the end of our conversion pass. A subset of such operations are shown in \cref{fig:identify-ops}. For each operation that has been identified, we provide a lowering that maps it to a legal operation in the \llvm dialect. We are able to use lowering patterns like \texttt{VectorConvertToLLVMPattern} from \mlir for common arithmetic and logical operations as shown in \cref{fig:convert-ops}.

\begin{figure*}[t]
\centering
\vfill
\begin{subfigure}[b]{0.5\textwidth}
\prettylstcpp
\begin{lstlisting}
void BtorToLLVMLoweringPass::runOnOperation() {
  LLVMConversionTarget target(getContext());
  RewritePatternSet patterns(&getContext());
  LLVMTypeConverter converter(&getContext());
  mlir::btor::populateBtorToLLVMConversionPatterns(converter, patterns);
  ...
  /// binary operators
  // arithmetic
  target.addIllegalOp<btor::AddOp, btor::SubOp, btor::MulOp, btor::UDivOp...>();
  ...
}
\end{lstlisting}
\caption{Identifying operations.}
\label{fig:identify-ops}
\end{subfigure}
\hspace{5mm}%
\hspace{5mm}%
\begin{subfigure}[b]{0.5\textwidth}
\prettylstcpp
\begin{lstlisting}
...
using AddOpLowering = VectorConvertToLLVMPattern<btor::AddOp, LLVM::AddOp>;
using SubOpLowering = VectorConvertToLLVMPattern<btor::SubOp, LLVM::SubOp>;
using MulOpLowering = VectorConvertToLLVMPattern<btor::MulOp, LLVM::MulOp>;
using UDivOpLowering = VectorConvertToLLVMPattern<btor::UDivOp, LLVM::UDivOp>;
...
void mlir::btor::populateBtorToLLVMConversionPatterns(
    LLVMTypeConverter &converter, RewritePatternSet &patterns) {
  patterns.add<
      AddOpLowering, SubOpLowering, MulOpLowering,
      UDivOpLowering, ...>(converter);
}
...
\end{lstlisting}
\caption{Converting operations to \llvmir.}
\label{fig:convert-ops}
\end{subfigure}
\caption[]{Using Patter Based Rewriters in \mlir.}
\label{fig:cpp}
\end{figure*}

We performed extensive testing using the HWMCC20 benchmark set to verify the correctness of \btormlir. This is the same benchmark set used to test ~\cite{btor2c}. The tests are run on a Linux machine with x86\_64 architecture, using \btormc with an unroll bound of~$20$, a timeout of $300$ seconds and memory limit of $65$ GB. 
 We present the results in \cref{tab:roundtrip}, where bitvector benchmarks categories are in the top half and bitvector + array benchmark categories are in the bottom half. All times in this table reflect solved instances and do not include timeouts. We do not show the time it takes to run \btormlir since the time is negligible. The results are grouped by competition contributor such that each row shows the time, instances solved (safe/unsafe) and timeouts (TO) for both the original and round-tripped circuits. For example, for the \texttt{wolf/18D} category, we can see that the original \btor circuit solves 34 safe instances and 0 unsafe instances in 157 seconds, with 2 timeouts. The round-tripped circuit solves 34 safe instances and 0 unsafe instances in 168 seconds with two timeouts. 
 
 We can see that the safety properties in \btor circuits are neither changed nor violated after being round-tripped by \btormlir. In two categories with only bitvectors, \texttt{19/beem} and \texttt{19/mann}, one more instance in each category is found safe after round trip, while the original circuit leads to a memout and timeout respectively. This gives us confidence that the translation to \btordialect, using the \btortools parser, is indeed correct. Then, we tested whether the same holds after translation to \llvmir. Through this method, we were able to ensure that \btormlir does not have errors when handling operations that are represented in the benchmark set. This approach is not complete, however, since it would not identify errors that might be in our implementation but are not exercised by the benchmarks we use. For example, \btor expects that a division by zero would result in $-1$, but there are no benchmarks that exercise this kind of division. We mitigate this by generating benchmarks for division, remainder, and modulus operators to ensure that the expected behavior of \btor operators are represented in our test suite. 

In the future, it is interesting to explore other translation validation and verification approaches. For example, it would be useful for \btormlir to produce a proof trail that justifies all of the transformations that are performed by the tool. This, for example, might be possible to achieve by building on the work of~\cite{DBLP:conf/dac/ChatterjeeMBK07,DBLP:conf/fmcad/Bryant22}.

\paragraph*{Limitations} \btormlir is able to round trip \btor operators and their sorts. In \llvmir all \btor operators and their sorts are supported as well, but not fairness and justice constraints. 

\section{Evaluation}
\label{sec:evaluation}

\begin{table*}[t]
\begin{minipage}{0.5\textwidth}
\centering
  \resizebox{1\textwidth}{!}{%
   \setlength{\tabcolsep}{4pt}
    \begin{tabular}{clrrrrrrr}
        & & \multicolumn{2}{c}{\btormc}  & & \multicolumn{3}{c}{\seahorn} \\[1mm]
        \cline{3-4}\cline{6-8}\\[-2mm]
         & & 20 & 25 & & VCGen + \zthree & VCGen & \textsc{Btor} \\[1mm]
         \toprule
         \multirow{4}{*}{\rotatebox[origin=c]{90}{\tt wolf/18D} \ } 
        & Time (s)\hspace{0.1in} & 157 &  394 & &  560 &  543 &  745 \\
        & Safe & 34 & 34 & & 29 & - & 34 \\
        & Unsafe & 0 & 0 & & 0 & - & 0 \\
        & TO & 2 & 2 & & 7 & 2 & 2 \\[1mm]
        \midrule
        \multirow{4}{*}{\rotatebox[origin=c]{90}{\tt wolf/19A} \ } 
        & Time (s) & 146 & 106 & &  - &  - & - \\
        & Safe & 0 & 0 & & 0 & - & 0 \\
        & Unsafe & 1 & 1 & & 0 & - & 0 \\
        & TO & 17 & 17 & & 18 & 18 & 18 \\[1mm]
        \midrule
        \multirow{4}{*}{\rotatebox[origin=c]{90}{\tt wolf/19B} \ } 
        & Time (s) & 2 & 2 & & 2 & 2 & 3 \\
        & Safe & 3 & 3 & & 3 & - & 3 \\
        & Unsafe & 0 & 0 & & 0 & - & 0 \\
        & TO & 0 & 0 & & 0 & 0 & 0 \\[1mm]
        \midrule
        \multirow{4}{*}{\rotatebox[origin=c]{90}{\tt wolf/19C} \ } 
        & Time (s) & 834 &  1\,101 & & 354 &  418 &   1\,085 \\
        & Safe & 108 & 107 & & 102 & - & 106 \\
        & Unsafe & 0 & 0 & &  0 & - & 0 \\
        & TO & 5 & 6 & & 11 & 2 & 7 \\[1mm]
        \bottomrule
        \vspace{0.1in}
    \end{tabular}
  }
\end{minipage}
\begin{minipage}{0.5\textwidth}
\centering
  \resizebox{1\textwidth}{!}{%
  \setlength{\tabcolsep}{4pt}
    \begin{tabular}{clrrrrrr}
        & & \multicolumn{2}{c}{\btormc} &  & \multicolumn{3}{c}{\seahorn} \\[1mm]
        \cline{3-4} \cline{6-8}\\[-2mm]
         & & 20 & 25 & & VCGen + \zthree & VCGen & \textsc{Btor} \\[1mm]
         \toprule
         \multirow{4}{*}{\rotatebox[origin=c]{90}{\tt 19/beem}\ } 
         & Time (s) & 278 & 251 & & 309 & 35 & 85 \\
         & Safe & 9 & 8 & & 6 & - & 7 \\
         & Unsafe & 2 & 2 & & 2 & - & 2 \\
         & TO & 4 & 5 & & 7 & 4 & 6 \\[1mm]
         \midrule
         \multirow{4}{*}{\rotatebox[origin=c]{90}{\tt 19/goel} \ } 
         & Time (s) & 190 & 349 &  & 489 & 132 & 335 \\
         & Safe & 26 & 25 & & 25 & - & 28 \\
         &Unsafe & 2 & 2 & & 1 & - & 2 \\
         & TO & 43 & 44 & & 45 & 27 & 41 \\[1mm]
         \midrule
         \multirow{4}{*}{\rotatebox[origin=c]{90}{\tt 19/mann} \ } 
         & Time (s)\hspace{0.05in} & 4\,442 & 8\,674 & & 3\,811 & 175 &  3\,015 \\
         & Safe & 29 & 28 & & 19 & - & 30 \\
         & Unsafe & 15 & 15 & & 14 & - & 14 \\
         & TO & 9 & 10 & & 20 & 2 & 9 \\[1mm]
         \midrule
         \multirow{4}{*}{\rotatebox[origin=c]{90}{\tt 20/mann} \ } 
        & Time (s) & 257 & 495 & & 94 & 35 &  188 \\
        & Safe & 10 & 10 & & 8 & - & 9 \\
        & Unsafe & 5 & 5 & & 5 & - & 5 \\
        & TO & 0 & 0 & & 2 & 0 & 1 \\[1mm]
        \bottomrule
        \vspace{0.1in}
    \end{tabular}
  }
\end{minipage}
\caption{HWMCC20 Results.}
\label{tab:results}
\end{table*}

\begin{figure*}[t]
  \centering
  \includegraphics[width=1\textwidth]{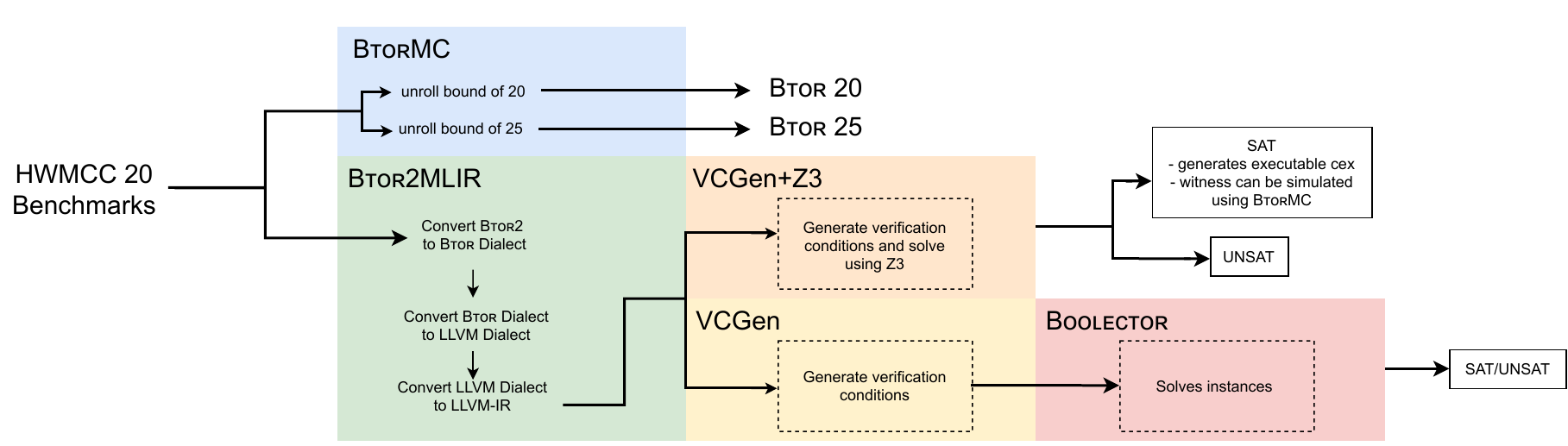}
  \caption{Verification Strategies.}
  \label{fig:strategies}
\end{figure*}

To evaluate \btormlir, we have built a prototype hardware model checker by connecting 
our tool with \seahorn~\cite{SeaHorn}, a well-known model checker for C/C++ programs that works at the LLVM-IR level. It has recently been extended with a bit-precise Bounded Model Checking engine~\cite{SeaBMC}. This BMC engine was evaluated in a recent case study~\cite{VVC} and we use the same configuration of \seahorn in our evaluation.

The goal of our evaluation is to show that \btormlir makes it easy to connect hardware designs with LLVM-based verification engines. We did not expect the existing software engines to outperform dedicated hardware model checkers. However, we hope that this will enable further avenues of research. In the future, we plan to extend the framework to support other LLVM-based analysis tools, such as symbolic execution engine KLEE~\cite{KLEE}, and fuzzing framework~\cite{LibFuzzer}.

For the evaluation, we have chosen the bitvector category of \textsc{btor} benchmarks from the most recent Hardware Model Checking Competition (HWMCC)~\cite{HWMCC}. We have excluded benchmarks with arrays since the export to \llvmir is not supported by \seahorn in our experimental setup. All our experiments are run on a Linux machine with x86\_64 architecture, with unroll bound of~$20$, a timeout of $300$ seconds and memory limit of $65$ GB. The results are presented in \cref{tab:results}, grouped by competition contributor. All times in this table reflect solved instances and do not include timeouts. We do not show the time it takes to run \btormlir since the time is negligible. In the rest of this section, we highlight some of the interesting findings.

We have run \btormc on the same machine and exact same experimental setup (unroll bound and CPU and memory limits). We chose \btormc because it is well integrated with the HWMCC environment and is specifically designed for \btor. The results of running \btormc are shown in the first columns of \btormc in \cref{tab:results}. For each category, we show the total time for all instances that are solved in that category, 
and the number of instances that are solved as safe, unsafe, and timed-out (TO), respectively. For example, the \texttt{20/mann} category is solved in 257 seconds, 10 instances are safe, 5 are unsafe, and no instance has timed out. The performance of \btormc is quite good across the board.

We evaluate the problems generated by \btormlir by plugging them into \seahorn. \seahorn pre-processes programs before attempting to verify them. This includes, standard \llvm optimizations (i.e., -O3), loop unrolling and loop cutting are applied. We found that \seahorn was able to, in some instances, remove the assertions in the \llvmir, meaning that the program was found to be safe statically, before invoking the BMC. The BMC also runs simplifications on the formulas that it sends to \zthree, its default underlying \smt solver. The results for this run are shown in the \zthree columns of \cref{tab:results}. For example, the \texttt{20/mann} category is solved in 94 seconds, 8 instances are safe, 5 are unsafe and 2 have timed out. The reported time does not include the instances that have timed out. 

The aggregate time of \seahorn on most of the categories is higher than that of \btormc, often by a significant amount. We looked into this and found that \seahorn treats the given bound as a lower bound, rather than an upper bound. That is, it ensures that it unrolls the programs to a depth of at least $20$, but it may continue past that point. Taking this into account, we ran \btormc with a bound of $25$. The results are in the second columns of \btormc in \cref{tab:results}. As expected, its aggregate times are higher than the run of \btormc with bound $20$. We notice, however, that it is slower than \seahorn in the \texttt{19/mann} category.

\boolector and \zthree are the \smt solvers used by \btormc and \seahorn respectively. Given that \boolector is optimized for \btor circuits, we evaluated whether the \smt formulas generated by \seahorn would be solved faster by \boolector. The results for generating \smtlib formulas using \seahorn are presented in the VCGen column of \cref{tab:results}. The times are low for most categories except \texttt{wolf/18D}, \texttt{wolf/19C}, \texttt{19/goel} and \texttt{19/mann}. For example, it takes \seahorn 175 seconds to generate the verification conditions for instances in the \texttt{19/mann} category, with two timeouts. This includes the time it takes \seahorn to print the \smt formulas to disk. We plug the resulting \smt formulas into \boolector and present the results in the \textsc{btor} columns of \cref{tab:results}. The results show that using \seahorn to generate verification conditions and \boolector to solve these instances is often better than using \btormc. For example, for category \texttt{19/mann}, it takes 3\,015s for \boolector to solve $44$ instances with $9$ timeouts. Therefore, the total time for \seahorn and \boolector (3\,190) represents the time it takes to translate, generate \smt formula and verify the \texttt{19/mann} category. Note that two of the 9 timeouts in this category are attributed to the fact that \seahorn has a timeout when generating verification conditions.

\begin{table}[t]
\setlength{\tabcolsep}{4pt}
    \centering
    \begin{tabular}{lrrrcrrr}
        & \multicolumn{2}{c}{\btormc} & & \multicolumn{3}{c}{\seahorn} \\
        \cline{2-3} \cline{5-7}
        \\[-2mm]
         & 20 & 25 & &VCGen+\zthree & VCGen & \textsc{Btor} \\
         \toprule
        Time (s)\hspace{0.1in} &  6\,309 &  11\,373 & &  5\,621 &  1\,340 &  5\,456 \\
        Safe & 219 &  215 & &  192 &  - &   217 \\
        Unsafe &  25 &  25 & & 22 &  - &  23 \\
        TO &  80 &   84 & & 110 &  55 &  84 \\
        \bottomrule
        \vspace{0.1in}
    \end{tabular}
    \caption{Total results for each tool.}
    \label{tab:total}
\end{table}

\begin{figure*}[t]
\centering
\begin{subfigure}[b]{0.32\textwidth}
    \includegraphics[width=1\textwidth]{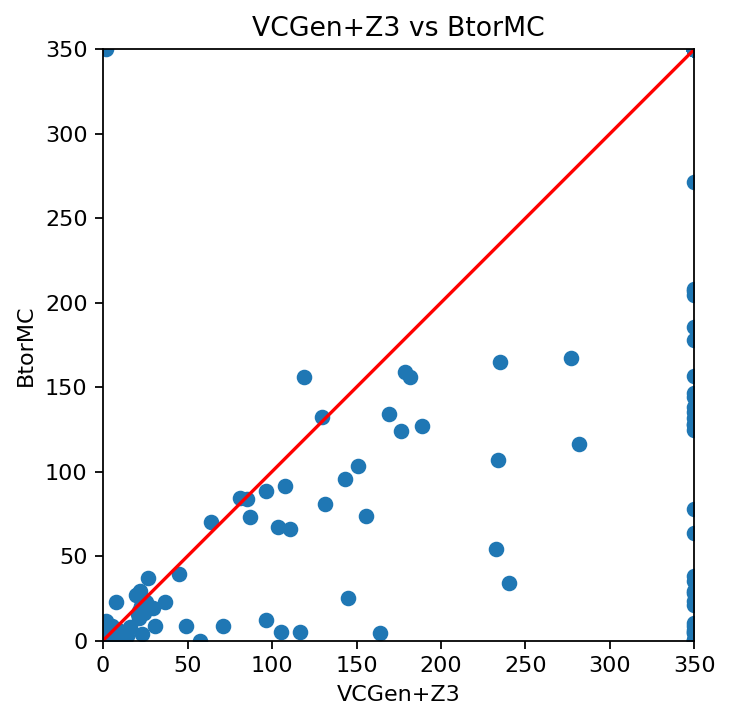}
    \vspace{0.2in}
\caption{VCGen + Z3 vs \btormc.}
\label{fig:bmcz3}
\end{subfigure}
\hfill%
\begin{subfigure}[b]{0.32\textwidth}
    \includegraphics[width=1\textwidth]{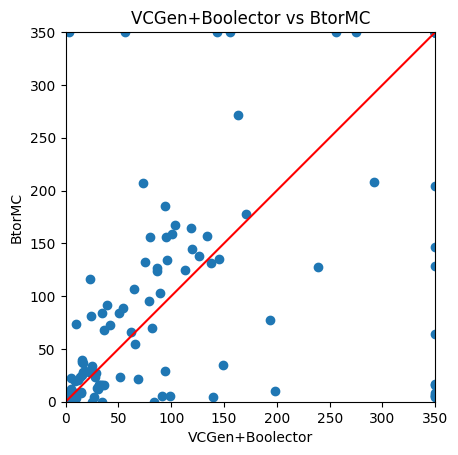}
    \vspace{0.2in}
\caption{VCGen + \boolector vs \btormc.}
\label{fig:bmcvc}
\end{subfigure}
\hspace*{\fill}%
\begin{subfigure}[b]{0.32\textwidth}
    \includegraphics[width=1\textwidth]{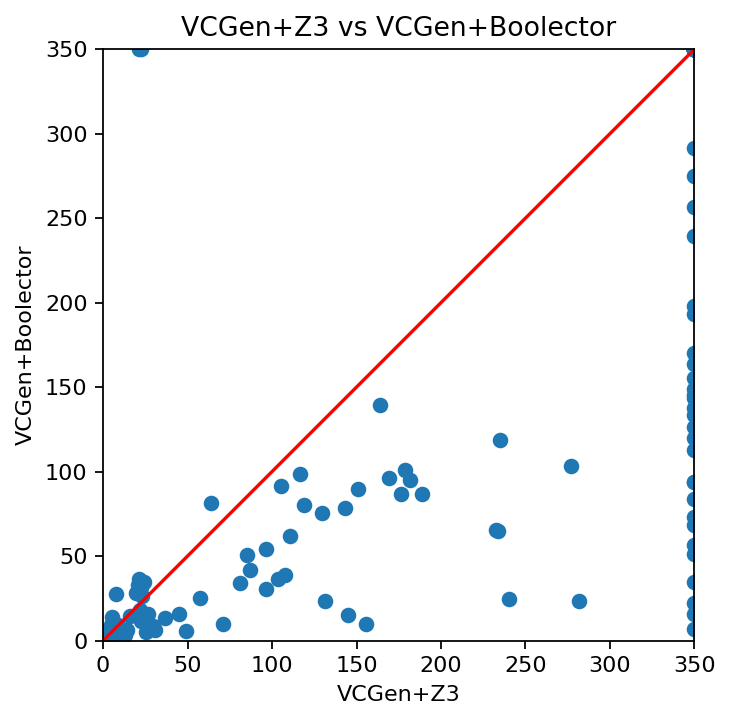}
    \vspace{0.2in}
\caption{VCGen + Z3 vs VCGen + \boolector.}
\label{fig:z3vc}
\end{subfigure}
\caption[]{Verification strategy comparison.}
\label{fig:plots}
\end{figure*}

To get the big picture of how the different infrastructures performed, we collected the results over all categories in \cref{tab:total}. From this table, we can see that our hybrid pipeline combining \btormlir, \seahorn, and \boolector solves $240$ instances with $84$ timeouts in $6\,796$s (sum of VCGen and \textsc{btor} total times), which is very encouraging. We also present plots that compare the different pipelines that have been explored in \cref{fig:plots}. We set the time for all timeout instances to 350 seconds so that they are distinguished from instances that were solved close to the timeout threshold. First, we look at the performance of the hybrid pipeline that combines \btormlir, \seahorn and its default \smt solver Z3 against \btormc in \cref{fig:bmcz3}. Z3 does as well as \btormc for most instances that are easy, however, it struggles when the problems are harder. This is not as clear from \cref{tab:results} since focuses on the number of timeouts and benchmarks solved. Second, we present the performance of \btormc against the hybrid pipeline that combines \btormlir, \seahorn and \boolector in \cref{fig:bmcvc}. We can see that there are more benchmarks that this pipeline solves faster than \btormc. It is also clear that it solves more benchmarks than the Z3 configuration in \cref{fig:bmcz3}, as we would expect from \cref{tab:total}. Third, we compare the two hybrid pipelines in \cref{fig:z3vc}. We can see that the configuration that uses \seahorn to generate verification conditions and \boolector for solving easily outperforms the Z3 configuration. 

\section{Related Work}
\label{sec:related}

Translating \btor circuits into other formats enables the application of different verification methods and techniques. The gains that can be made from applying one method of encoding over another could enable solving a class of benchmarks that are not  solved with existing approaches. 

\textsc{Btor2LLVM}\footnote{\url{https://github.com/stepwise-alan/btor2llvm}} and \textsc{Btor2CHC}\footnote{\url{https://github.com/stepwise-alan/btor2chc}} are tools that convert \btor circuits to programs in \llvmir and CHCs, respectively. These tools are developed in Python, in order to be light weight, but end up repeating shared functionality and tools since they lack a common infrastructure. Translated \btor benchmarks\footnote{\url{https://github.com/zhanghongce/HWMCC19-in-CHC}} have also been collected to facilitate research, but information of what tools were used to get the CHC format is not publicly available. While a collection of translated benchmarks is valuable, it is important that there are tools to do the translation on demand. This enables rapid prototyping in a way that saved benchmarks do not. 

\textsc{Btor2C}~\cite{btor2c} is a recent tool that converts \btor circuits to C programs. It has been used to facilitate the utilization of software analyzers by serving as a pre-processing step that bridges the gap between the world of software verification and hardware verification. There are limitations that arise, however, from differences in the semantics of \btor and C. An important limitation that C imposes on this project is the inability to represent arbitrary width bitvectors. This means that \btor circuits which operate on bitvectors of width greater than 128 are not supported. These limitations, as well as \textsc{Btor2C} lack of support for \btor operators that have overflow detection are resolved by using \llvmir as the target language. 

A common theme across these efforts is that they are not built on an architecture that can be easily extended. Each project aims to make it easier to utilize advances in formal verification, but they fail to offer a solution that does not require recreating components that already exist. 

\section{Conclusion}
\label{sec:conclusion}

In this paper, we present \btormlir\ --- a new format and toolchain for hardware verification, based on the \mlir intermediate representation framework of the \llvm compiler infrastructure. Our goal is to open new doors for the research and applications of hardware verification by taking advantage of recent innovations in compiler construction technology. We believe that this project opens new avenues for exploring the application of existing verification and testing techniques developed for software to the hardware domain. As a proof of concept, we have connected \btormlir with the \textsc{SeaHorn} verification engine. While out-of-the-box, this gives acceptable performance, when combined with \textsc{Boolector}, a combination that is competitive against \textsc{BtorMC}. In the future, we plan to continue this line of research and explore applying testing and simulation technologies such as \klee~\cite{KLEE} and \libfuzzer~\cite{LibFuzzer}. We also plan to generate formats for other verification techniques such as \textsc{Aiger}~\cite{aiger}, Constrained Horn Clauses, and \smtlib.

%
%
%
\bibliographystyle{IEEEtran}
\bibliography{bibliography}

\begin{thebibliography}{10}
\providecommand{\url}[1]{#1}
\csname url@samestyle\endcsname
\providecommand{\newblock}{\relax}
\providecommand{\bibinfo}[2]{#2}
\providecommand{\BIBentrySTDinterwordspacing}{\spaceskip=0pt\relax}
\providecommand{\BIBentryALTinterwordstretchfactor}{4}
\providecommand{\BIBentryALTinterwordspacing}{\spaceskip=\fontdimen2\font plus
\BIBentryALTinterwordstretchfactor\fontdimen3\font minus
  \fontdimen4\font\relax}
\providecommand{\BIBforeignlanguage}[2]{{%
\expandafter\ifx\csname l@#1\endcsname\relax
\typeout{** WARNING: IEEEtran.bst: No hyphenation pattern has been}%
\typeout{** loaded for the language `#1'. Using the pattern for}%
\typeout{** the default language instead.}%
\else
\language=\csname l@#1\endcsname
\fi
#2}}
\providecommand{\BIBdecl}{\relax}
\BIBdecl

\bibitem{HV}
S.~Malik, ``Hardware verification: Techniques, methodology and solutions,'' in
  \emph{Tools and Algorithms for the Construction and Analysis of Systems},
  C.~R. Ramakrishnan and J.~Rehof, Eds.\hskip 1em plus 0.5em minus 0.4em\relax
  Berlin, Heidelberg: Springer Berlin Heidelberg, 2008, pp. 1--1.

\bibitem{HWMCC}
A.~Biere, T.~van Dijk, and K.~Heljanko, ``Hardware model checking competition
  2017,'' in \emph{2017 Formal Methods in Computer Aided Design (FMCAD)}, 2017,
  pp. 9--9.

\bibitem{BTOR}
A.~Niemetz, M.~Preiner, C.~Wolf, and A.~Biere, ``{Btor2 , BtorMC and Boolector
  3.0},'' in \emph{Computer Aided Verification}, H.~Chockler and
  G.~Weissenbacher, Eds.\hskip 1em plus 0.5em minus 0.4em\relax Cham: Springer
  International Publishing, 2018, pp. 587--595.

\bibitem{LLVM}
C.~Lattner and V.~Adve, ``{LLVM: a compilation framework for lifelong program
  analysis \& transformation},'' in \emph{International Symposium on Code
  Generation and Optimization, 2004. CGO 2004.}, 2004, pp. 75--86.

\bibitem{MLIR}
C.~Lattner, M.~Amini, U.~Bondhugula, A.~Cohen, A.~Davis, J.~Pienaar, R.~Riddle,
  T.~Shpeisman, N.~Vasilache, and O.~Zinenko, ``{MLIR: A Compiler
  Infrastructure for the End of Moore's Law},'' 2020.

\bibitem{LibFuzzer}
K.~Serebryany, ``{Continuous Fuzzing with libFuzzer and AddressSanitizer},'' in
  \emph{2016 IEEE Cybersecurity Development (SecDev)}, 2016, pp. 157--157.

\bibitem{KLEE}
C.~Cadar, D.~Dunbar, and D.~R. Engler, ``Klee: Unassisted and automatic
  generation of high-coverage tests for complex systems programs,'' in
  \emph{USENIX Symposium on Operating Systems Design and Implementation}, 2008.

\bibitem{SeaHorn}
A.~Gurfinkel, T.~Kahsai, A.~Komuravelli, and J.~A. Navas, ``{The SeaHorn
  Verification Framework},'' in \emph{Computer Aided Verification}, D.~Kroening
  and C.~S. P{\u{a}}s{\u{a}}reanu, Eds.\hskip 1em plus 0.5em minus 0.4em\relax
  Cham: Springer International Publishing, 2015, pp. 343--361.

\bibitem{DBLP:conf/vstte/GurfinkelN21}
\BIBentryALTinterwordspacing
A.~Gurfinkel and J.~A. Navas, ``Abstract interpretation of {LLVM} with a
  region-based memory model,'' in \emph{Software Verification - 13th
  International Conference, {VSTTE} 2021, New Haven, CT, USA, October 18-19,
  2021, and 14th International Workshop, {NSV} 2021, Los Angeles, CA, USA, July
  18-19, 2021, Revised Selected Papers}, ser. Lecture Notes in Computer
  Science, R.~Bloem, R.~Dimitrova, C.~Fan, and N.~Sharygina, Eds., vol.
  13124.\hskip 1em plus 0.5em minus 0.4em\relax Springer, 2021, pp. 122--144.
  [Online]. Available: \url{https://doi.org/10.1007/978-3-030-95561-8\_8}
\BIBentrySTDinterwordspacing

\bibitem{CIRCT}
S.~Eldridge, P.~Barua, A.~Chapyzhenka, A.~Izraelevitz, J.~Koenig, C.~Lattner,
  A.~Lenharth, G.~Leontiev, F.~Schuiki, R.~Sunder, A.~Young, and R.~Xia,
  ``{MLIR as Hardware Compiler Infrastructure},'' in \emph{Workshop on
  Open-Source EDA Technology (WOSET)}, 2021.

\bibitem{SMTLIB}
C.~Barrett, P.~Fontaine, and C.~Tinelli, ``{The Satisfiability Modulo Theories
  Library (SMT-LIB)},'' {\tt www.SMT-LIB.org}, 2016.

\bibitem{Yosys}
C.~Wolf, ``Yosys open synthesis suite,'' \url{https://yosyshq.net/yosys/}.

\bibitem{Verilog}
S.~Palnitkar, \emph{Verilog HDL: A Guide to Digital Design and
  Synthesis}.\hskip 1em plus 0.5em minus 0.4em\relax USA: Prentice-Hall, Inc.,
  1996.

\bibitem{TensorFlow}
\BIBentryALTinterwordspacing
M.~Abadi, P.~Barham, J.~Chen, Z.~Chen, A.~Davis, J.~Dean, M.~Devin,
  S.~Ghemawat, G.~Irving, M.~Isard, M.~Kudlur, J.~Levenberg, R.~Monga,
  S.~Moore, D.~G. Murray, B.~Steiner, P.~A. Tucker, V.~Vasudevan, P.~Warden,
  M.~Wicke, Y.~Yu, and X.~Zheng, ``{TensorFlow}: {A} system for large-scale
  machine learning,'' in \emph{12th {USENIX} Symposium on Operating Systems
  Design and Implementation, {OSDI} 2016, Savannah, GA, USA, November 2-4,
  2016}, K.~Keeton and T.~Roscoe, Eds.\hskip 1em plus 0.5em minus 0.4em\relax
  {USENIX} Association, 2016, pp. 265--283. [Online]. Available:
  \url{https://www.usenix.org/conference/osdi16/technical-sessions/presentation/abadi}
\BIBentrySTDinterwordspacing

\bibitem{btor2c}
\BIBentryALTinterwordspacing
D.~Beyer, P.-C. Chien, and N.-Z. Lee, ``Bridging hardware and software analysis
  with {Btor2C}: {A} word-level-circuit-to-{C} translator,'' in \emph{Proc.\
  TACAS}, ser. LNCS~13994.\hskip 1em plus 0.5em minus 0.4em\relax Springer,
  2023, pp. 1--21. [Online]. Available:
  \url{https://www.sosy-lab.org/research/btor2c/}
\BIBentrySTDinterwordspacing

\bibitem{DBLP:conf/dac/ChatterjeeMBK07}
\BIBentryALTinterwordspacing
S.~Chatterjee, A.~Mishchenko, R.~K. Brayton, and A.~Kuehlmann, ``On resolution
  proofs for combinational equivalence,'' in \emph{Proceedings of the 44th
  Design Automation Conference, {DAC} 2007, San Diego, CA, USA, June 4-8,
  2007}.\hskip 1em plus 0.5em minus 0.4em\relax {IEEE}, 2007, pp. 600--605.
  [Online]. Available: \url{https://doi.org/10.1145/1278480.1278631}
\BIBentrySTDinterwordspacing

\bibitem{DBLP:conf/fmcad/Bryant22}
\BIBentryALTinterwordspacing
R.~E. Bryant, ``Tbuddy: {A} proof-generating {BDD} package,'' in \emph{22nd
  Formal Methods in Computer-Aided Design, {FMCAD} 2022, Trento, Italy, October
  17-21, 2022}, A.~Griggio and N.~Rungta, Eds.\hskip 1em plus 0.5em minus
  0.4em\relax {IEEE}, 2022, pp. 49--58. [Online]. Available:
  \url{https://doi.org/10.34727/2022/isbn.978-3-85448-053-2\_10}
\BIBentrySTDinterwordspacing

\bibitem{SeaBMC}
S.~Priya, X.~Zhou, Y.~Su, Y.~Vizel, Y.~Bao, and A.~Gurfinkel, ``{Bounded Model
  Checking for LLVM},'' in \emph{Formal Methods in Computer Aided Design,
  {FMCAD} 2022}, 2022, p. 214.

\bibitem{VVC}
\BIBentryALTinterwordspacing
------, ``Verifying verified code,'' \emph{Innov. Syst. Softw. Eng.}, vol.~18,
  no.~3, pp. 335--346, 2022. [Online]. Available:
  \url{https://doi.org/10.1007/s11334-022-00443-9}
\BIBentrySTDinterwordspacing

\bibitem{aiger}
A.~Biere, K.~Heljanko, and S.~Wieringa, ``{AIGER 1.9} and beyond,'' Institute
  for Formal Models and Verification, Johannes Kepler University,
  Altenbergerstr. 69, 4040 Linz, Austria, Tech. Rep. 11/2, 2011.

\end{thebibliography}

\end{document}